\documentstyle[12pt,moriond,times,epsfig]{article}
\begin{document}
%Here you should enter the title of your manuscript
\heading{THE CHEMICAL EVOLUTION OF GALAXIES BY SUCCESSIVE STARBURSTS}

\author{
T.\ Contini $^{1}$, 
R.\ Coziol $^{2}$, 
S.\ Consid\`ere $^{3}$, 
E.\ Davoust $^{4}$, 
\& R.E.\ Carlos Reyes $^{5}$}
{$^{1}$ European Southern Observatory, Garching bei M\"unchen, Germany}  
{$^{2}$ Osservatorio Astronomico di Brera, Milano, Italy}
{$^{3}$ Observatoire de Besan\c{c}on, Besan\c{c}on, France}
{$^{4}$ Observatoire Midi-Pyr\'en\'ees, Toulouse, France}
{$^{5}$ Seminario Permanente de Astronom\'{\i}a y Ciencas Espaciales, L\'{\i}ma, Per\'u}

\begin{moriondabstract}

We propose an evolutionary scenario by successive bursts of star formation 
to reproduce the chemical properties of massive nearby Starburst Nucleus
Galaxies (SBNGs). The N/O abundance ratios in SBNGs are $\sim$ 0.2 dex higher
than in normal H\,{\sc ii} regions observed in the disks of
late--type spirals.
%, but comparable to the values found in the bulges
%of normal early--type spirals.  
The variation of the N/O ratio as a
function of metallicity follows a $primary + secondary$ relation, but
the increase of nitrogen does not appear as a continuous process.
Assuming that nitrogen is produced by intermediate-mass stars, we
show that our observations are consistent with a model where the bulk
of nitrogen in SBNGs was formed during past sequences of bursts of star
formation which probably started 2 or 3 Gyrs in the past. 
%What we observe, therefore, could be the main production of nitrogen in
%the bulges of these galaxies.

\end{moriondabstract}

\vspace{-.5cm}
\section{Introduction}

Recent observations obtained with the {\em Hubble Space Telescope} have made
clear the urgency 
of understanding the nature of the starburst phenomenon.
Drastic and rapid changes of the population of galaxies have been observed
over a short period of time and at a surprisingly recent epoch 
(\cite{vdBetal96},\cite{Detal98})
%,\cite{CB99})
.  These
observations seem to imply that most galaxies formed between redshifts of
1 and 2 (\cite{MPD98}) and support the hierarchical formation of galaxies
paradigm.  According to this theory, massive galaxies form by successive
mergers of smaller mass and gas rich components.  If each of these mergers
triggers a burst of star formation (\cite{TL79}), then,
consequently, the galaxies form and evolve by a succession of bursts.  Is
this what we observe in present-day starburst galaxies?  The subsequent
discovery of forming galaxies at high redshifts with spectral characteristics
in the UV similar to those of nearby starbursts 
(\cite{Setal96},\cite{Setal99}) supports such an interpretation.

That many nearby SBNGs could be the remnants of merging galaxies is already
suggested by several observations 
(\cite{KvS92},\cite{CBD95},\cite{BCD95}). 
It has also been shown that {\em i)} SBNGs are
chemically less evolved than normal galaxies with similar morphologies and
comparable luminosities (\cite{Cozetal97a}), {\em ii)} they are
predominantly early--type spirals (\cite{Cozetal97b},\cite{Cozetal98a}) 
and {\em
iii)} they follow a luminosity--metallicity relation similar to that of
elliptical galaxies (\cite{Cozetal98b}).  The SBNGs also have another
intriguing property which makes them similar to star-forming galaxies at
high redshifts.  In their analysis of the properties of the Lyman-break 
galaxies, \cite{Setal96} concluded that the star formation rate
in these galaxies was probably constant over the last Gyr.  In the case of
SBNGs, it has been demonstrated that multiple bursts of star formation
over a few Gyr period produce nearly constant star formation rates in these 
galaxies (
%\cite{CD95}
,\cite{Coz96},\cite{DCD99}).
But are these sequences of bursts the consequence of multiple merger events?

In order to gain new insights on the nature and origin of the nearby
SBNGs, we have embarked in a new project to establish a more complete
picture of their chemical evolution.  
A new method has recently been devised for estimating
nitrogen abundances in metal-rich galaxies (\cite{TEH96}).  
We have taken advantage
of this important advance to determine the abundance of nitrogen in
SBNGs and compare it with the values observed in normal spirals.

\vspace{-.5cm}
\section{The abundance of nitrogen in SBNGs}

Our sample of SBNGs was composed originally of 208 H\,{\sc ii}
regions observed along the bars of 75 Markarian barred galaxies (see
\cite{Cont96},\cite{CCD98},\cite{Consetal99} for details). 
From this sample, we rejected 48 H\,{\sc ii} regions
because of their ambiguous classification using three spectroscopic
diagnostic diagrams.  The same criterion was used by \cite{Vetal95} 
to build their sample of 83 FIR-bright SBNGs.  After verifying
that they have similar spectroscopic characteristics, we merged the two
samples together. The detailed analysis can be found in \cite{Cozetal99}.

In SBNGs, nitrogen appears overabundant as compared to ``normal'' 
disk H\,{\sc ii} regions, with a relative abundance N/O which is 
$\sim$ 0.2 dex higher (\cite{Cozetal99}). 
The range of N/O values found in SBNGs is comparable to that
observed in the bulges of normal early-type spiral galaxies 
(\cite{TEH96},\cite{vZSH98}). 
On this matter, our observations are consistent
with the recent discovery made by \cite{TEH96}, 
who showed that H\,{\sc
ii} regions in early-type spirals have slightly higher N/O ratios
than H\,{\sc ii} regions in late-type spirals.  
%%Because H\,{\sc ii}
%%regions in late-type spirals are generally more numerous and luminous
%%than in early-type spirals, samples of H\,{\sc ii} regions in normal
%%galaxies are naturally biased towards those in late-type spirals.
In normal galaxies,  samples of H\,{\sc ii} regions are preferentially
found in late-type spirals where they are generally more numerous and 
luminous than in early-type galaxies.
The SBNGs, on the other hand, are more numerous among
early-type spirals (\cite{Cozetal98b}), explaining the
observed higher abundance ratio in this sample.

We conclude that our measurements of the nitrogen abundance in SBNGs
are consistent with the chemical evolution of early-type spiral
galaxies.  This suggests that what we see could be the main production
of nitrogen in the bulges of these galaxies.

\vspace{-.5cm}
\section{An evolution by successive starbursts}

Examining how the abundance of nitrogen varies with the increase of oxygen,
we find that, contrary to normal disk H\,{\sc ii} regions, the SBNGs
do not follow the $secondary$ relation (see Fig.\ 1a).  
A linear fit yields (with a correlation coefficient of 76\%) 
$\log($N/O$) = 0.55\log($O/H$) + 0.8$, which is consistent with a 
mixture of $primary + secondary$
mode of production of nitrogen.
% (\cite{McG91}).
  But the increase of the N/O
ratio with metallicity does not seem to follow a
continuous process.  The N/O ratio rises sharply by about 0.3 dex at
an oxygen abundance of $\sim -3.4$ and stays almost constant in the
range $-3.4 <$log(O/H)$< - 2.9$.  
%Because of the high number of
%points (243) in this figure, the lack of values on the $secondary$
%relation over this particular range of metallicities cannot be
%attributed to an incompleteness of the data.

\begin{figure}[t] 
\hskip -4cm \epsfig{file=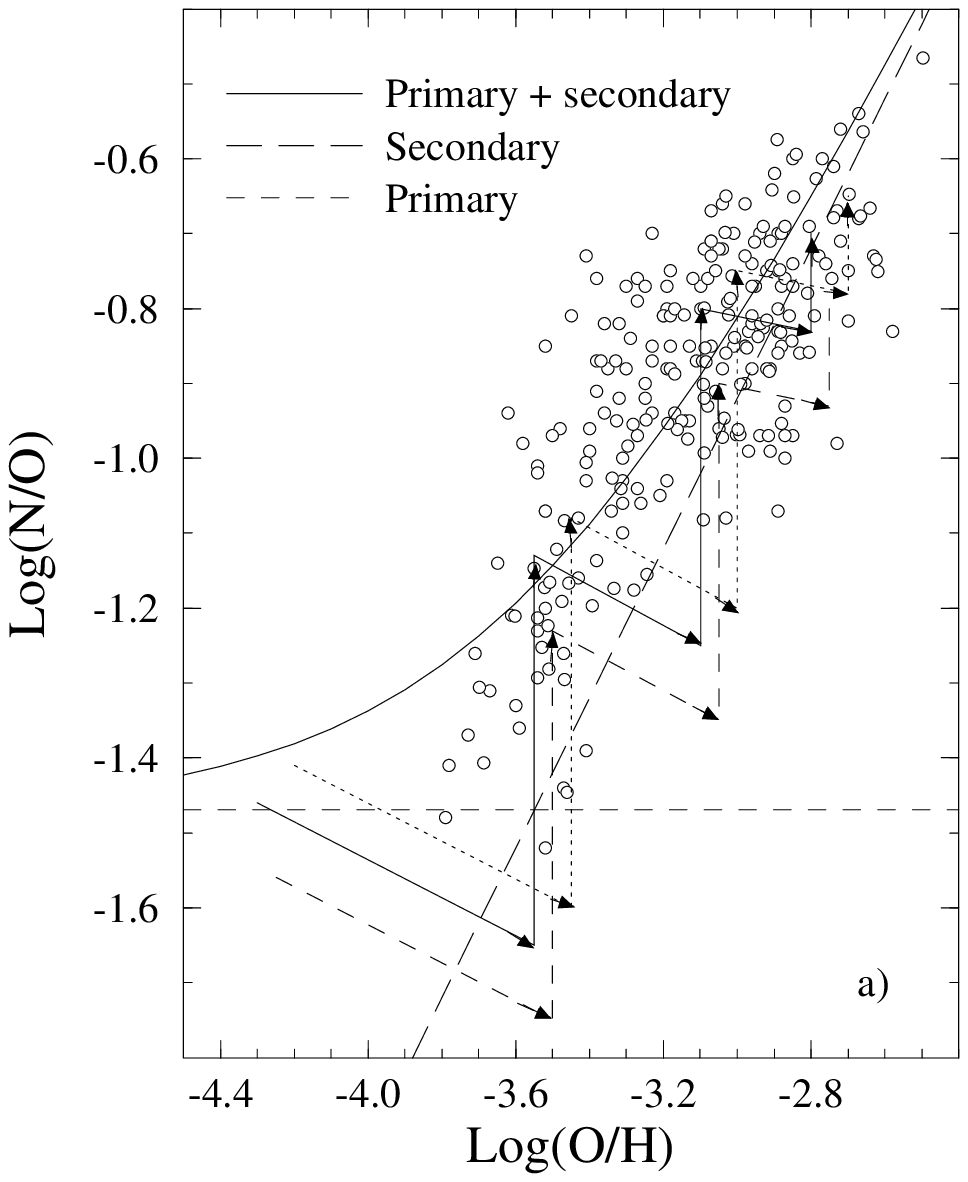,width=30pc}
\hskip -5cm \epsfig{file=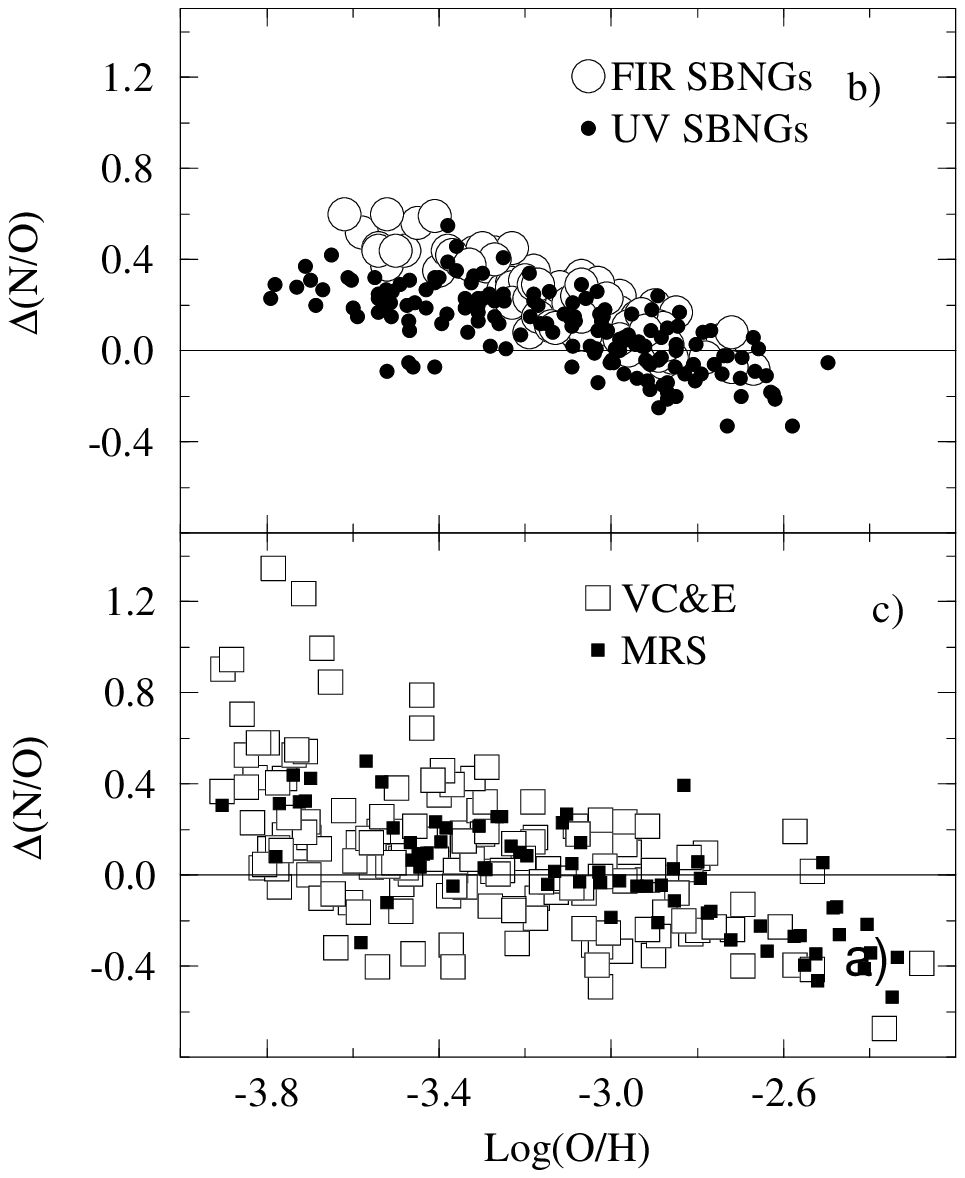,width=30pc} \vskip -8.5cm
\caption{\small 
a) Schematic representation of the process of production of
nitrogen in SBNGs by a sequence of bursts.  The dispersion is caused by
different initial intensities or different ages of the bursts
(represented by only three different vector sums in this figure).
Deviation of the observed nitrogen abundance from the $secondary$
relation ($\Delta$(N/O)) vs.  oxygen abundance:  b) in the SBNGS, and
c) in normal H\,{\sc ii} regions.  
Note how the behavior becomes more starburst-like for normal
galaxies with low metallicities.  } 
\end{figure}

In Figure~1a, we show schematically how a sequence of bursts could
explain our observations.  Our scenario is based on the analytical
model presented in \cite{G90}.  We assume that SBNGs begin their
chemical evolution with N/O and O/H ratios typical of H\,{\sc ii}
galaxies.  Massive stars are responsible for the increase in oxygen,
while nitrogen is only the product of intermediate-mass stars (\cite{vdHG97}).

During the first burst, the rapid evolution of massive stars increases
O/H and decreases N/O (\cite{G90},\cite{O95}).  Then, after
$\sim 0.4$ Gyr, the evolution of intermediate--mass stars increases
only N/O.  Models of sequential bursts usually predict that
successive bursts will have decreasing intensities (\cite{GSS80},\cite{KT93}). 
%,\cite{MMT94},\cite{KTH95}).  
A second burst, therefore, will produce a slightly lower
increase of O/H.  The decrease in N/O during oxygen enrichment (the
slope of the vector) will also be smaller as it becomes more and more
difficult to lower this ratio when the oxygen abundance increases
(\cite{G90}). Again, 0.4 Gyr after the beginning of the second
burst, N/O will increase, but with an amplitude relatively
smaller than in the first burst.  If we increase the number of
bursts and assume that successive bursts get weaker and weaker, the
sum of the vectors should converge towards a line whose slope represents
the mean increase of O/H and N/O in time.  This slope may resemble
the $secondary$ relation. Indeed, it is interesting to note that, according
to this scenario, a constant star formation is similar to an
infinite sum of very low-intensity bursts of star formation, which is
also consistent with the behavior of normal disk H\,{\sc ii} regions.

The above model predicts that the deviation of the observed N/O ratio from
the $secondary$ relation ($\Delta$(N/O)) will decrease with the age of
starburst galaxies, or, equivalently, with their increase in metallicity.  In
Figure~1b, we see that this prediction is satisfied in the SBNGs.  The fact
that the deviation in Figure~1b is mostly positive could be explained by the
different durations of the chemical evolutionary phases.  Because massive
stars have very short lifetimes, the oxygen enrichment phase is almost
instantaneous, and O/H rapidly reaches a maximum (the tip of each horizontal
vector in Figure~1a).  The lifetime of the stars producing nitrogen, on the
other hand, spans a much larger range of values.  The nitrogen enrichment
probably increases rapidly at the beginning, but extends over a longer period
of time, as lower and lower-mass stars evolve.  As a result, the top of each
vertical vector in Figure~1a is always much more populated, this phase
representing a sort of natural stable mode in the starburst's evolution.

According to the model of \cite{CB91}, the main phase of
production of nitrogen in a burst should occur between 0.4 and 1.6 Gyrs after
its beginning (this implies the evolution of stars from 8 to 3
M$_\odot$).  Considering nearly constant star formation rates over the last
2--3 Gyr (\cite{Coz96}), and assuming a median age of 1 Gyr for one burst, it
seems therefore that 2 or 3 bursts (or more if the bursts have shorter
durations) are necessary to produce a sufficient number of co-evolved
intermediate-mass stars needed to produce in turn
the observed abundance of nitrogen.
This would then push the origin of the main bursts 2 to 3 Gyrs in the past.

\section{Conclusion}

\vspace{-.3cm}
The time scales deduced above for the origin of the bursts in SBNGs are much
longer than those predicted by interacting-merging galaxy models (\cite{MH94},
\cite{MH96} for example).  It is not clear, therefore, what
internal/external phenomenon could allow SBNGs to form stars over such a long
period of time.  In the literature, we know of only two models which predict
sequences of bursts of star formation.  The ``stochastic self propagation of
star formation'' theory (\cite{GSS80},\cite{KT93})
and the ``hierarchical formation of galaxies'' theory (\cite{TL79}).

It is generally accepted that the Lyman-break galaxies are the progenitors of
present-day normal massive galaxies (\cite{Setal96}). The fact that
the SBNGs show similar characteristics to those of
these galaxies suggests, therefore,
that they may be nearby examples of galaxies still in formation.  The
SBNGs, consequently, would not be a peculiar phase in the evolution of
galaxies, but the result of a process which was much more common in the past
of the Universe.  This process could be the hierarchical formation of 
galaxies.

\vfill
\end{document}